\documentclass[fleqn,usenatbib]{mnras}

\usepackage{newtxtext,newtxmath}

\usepackage[T1]{fontenc}

\DeclareRobustCommand{\VAN}[3]{#2}
\let\VANthebibliography\thebibliography
\def\thebibliography{\DeclareRobustCommand{\VAN}[3]{##3}\VANthebibliography}

\usepackage{graphicx}	
\usepackage{amsmath}	
\usepackage{enumitem}

\title[Accreted stellar kinematics and DM halo spin]{A correlation between accreted stellar kinematics and dark matter halo spin in the \textsc{artemis} simulations}

\author[A. M. Dillamore et al.]{
Adam M. Dillamore,$^{1}$\thanks{E-mail: amd206@cam.ac.uk}
Vasily Belokurov,$^{1,2}$
N. Wyn Evans$^{1}$
and Andreea S. Font$^{3}$
\\
$^{1}$Institute of Astronomy, University of Cambridge, Madingley Road, Cambridge CB3 0HA, UK\\
$^{2}$Center for Computational Astrophysics, Flatiron Institute, 162 5th Avenue, New York, NY 10010, USA\\
$^{3}$Astrophysics Research Institute, Liverpool John Moores University, 146 Brownlow Hill, Liverpool L53RF, UK
}

\date{Accepted XXX. Received YYY; in original form ZZZ}

\pubyear{2022}

\begin{document}
\label{firstpage}
\pagerange{\pageref{firstpage}--\pageref{lastpage}}
\maketitle

\begin{abstract}
We report a correlation between the presence of a \textit{Gaia}-Sausage-Enceladus (GSE) analogue and dark matter halo spin in the \textsc{artemis} simulations of Milky Way-like galaxies. The haloes which contain a large population of accreted stars on highly radial orbits (like the GSE) have lower spin on average than their counterparts with more isotropic stellar velocity distributions. The median modified spin parameters $\lambda^\prime$ differ by a factor of $\sim1.7$ at the present-day, with a similar value when the haloes far from virial equilibrium are removed. We also show that accreted stars make up a smaller proportion of the stellar populations in haloes containing a GSE analogue, and are stripped from satellites with stellar masses typically $\sim4$ times smaller. Our findings suggest that the higher spin of DM haloes without a GSE-like feature is due to mergers with large satellites of stellar mass $\sim10^{10}M_\odot$, which do not result in prominent radially anisotropic features like the GSE.
\end{abstract}




\section{Introduction}
Angular momentum is one of the fundamental properties of dark matter (DM) haloes. It is traditionally quantified by the dimensionless spin parameter \citep{Peebles_71}, defined as
\begin{equation}\label{equation:lambda}
    \lambda \equiv \frac{J|E|^{1/2}}{GM^{5/2}},
\end{equation}
where J, E and M are the total angular momentum, energy and mass of the halo. A halo's spin results from a combination of tidal torques from the surrounding mass distribution, and accumulation of angular momentum from accreted matter. Tidal torque theory \citep{Hoyle_1951, Peebles_69, Doroshkevich, White_84} describes how haloes gain their initial angular momentum via tidal torques from neighbouring overdensities. These arise due to misalignment between the halo's inertia tensor and the gravitational tidal tensor.

Numerical simulations predict that the distribution of $\lambda$ (or the modified spin parameter $\lambda^\prime$, see Section~\ref{section:calculation}) across many haloes is roughly log-normal in shape, with a median of $\lambda\sim0.05$ and standard deviation in $\mathrm{ln}\lambda$ of $\sigma\sim0.5$ \citep{Barnes,Bullock,Gardner,Bailin,Munoz}.

The spin of DM haloes also evolves due to accretion and merger events. Major merger events tend to increase the average spin \citep{Gardner,Maller} due to acquisition of the accreted satellites' angular momentum. \citet{Vitvitska} found that a halo's spin usually sharply increases during a major merger, but gradually decreases while small satellites are accreted. This can be modelled as a random walk. It is therefore expected that two samples of galaxies with different accretion histories would have distinct spin distributions. However, \citet{D'Onghia} argued that this difference is due to recently merged haloes being out of virial equilibrium. The surrounding large scale structure (LSS) of a halo can also affect its spin. For example, the Milky Way (MW) lies in a structure known as the `Local Sheet' \citep{Tully}. \citet{Aragon-Calvo} found that simulated haloes located in such sheets tend to have lower spins, with strong alignment between the sheets and the haloes' angular momentum. \citet{Obreja} recently compared the angular momentum of DM and stellar components of haloes in simulations, and derived a relation between them. This was used to predict the spin of the Milky Way's DM halo, although accretion history was not directly considered in this calculation.

The \textit{Gaia} Sausage-Enceladus (GSE) is a population of comparatively metal-rich stars on highly eccentric orbits in the stellar halo of the Milky Way (MW). It was discovered by \citet{Belokurov_sausage}, and confirmed by \citet{Helmi_enceladus} using data from the second data release (DR2) of the \textit{Gaia} space telescope \citep{Gaia_16}. Comparisons with cosmological simulations have shown that these stars must have originated in a massive satellite of total mass $\sim10^{11}M_\odot$ which merged with the MW 8-11 Gyr ago \citep[e.g.][]{Belokurov_sausage, Fattahi}. This merger had a transformative impact on the MW, ejecting stars from the disc into the stellar halo \citep[the `Splash'; see][]{Bonaca_splash, Belokurov_splash}. Due to the merger's high total mass ratio of around 1:2.5 \citep{Naidu_simulations}, it likely had a large impact on the DM halo of the MW, such as reshaping or reorientation \citep{Dillamore}. It is also reasonable to expect that the halo's spin was affected. In this \textit{Letter}, we investigate whether the presence or absence of a GSE-like feature in a MW-like galaxy correlates with its DM halo's spin. We use the \textsc{artemis} set of 45 high resolution zoom-in simulations of MW-mass galaxies \citep{Font}, which have already been studied in relation to the GSE's impact on the MW \citep{Dillamore}.

The \textit{Letter} is arranged as follows. In Section~\ref{section:simulations}, we briefly describe the \textsc{artemis} simulations used in this study. We explain the methods for selecting halo samples and calculating spin parameters in Section~\ref{section:method} before presenting our results in Section~\ref{section:results}. Finally, our findings are summarised and discussed in Section~\ref{section:summary}.

\begin{figure}
  \centering
  \includegraphics[width=0.9\columnwidth]{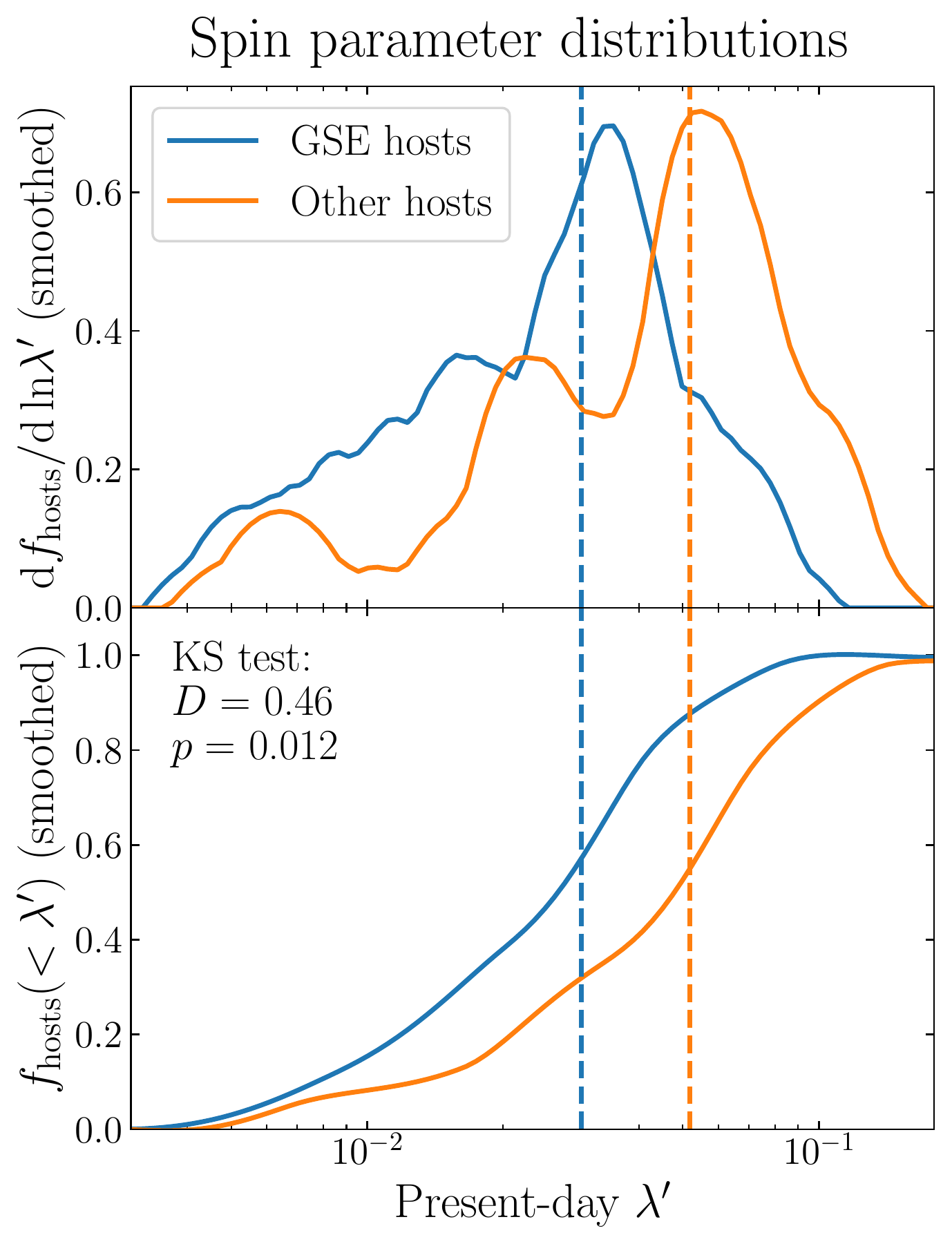}
  \caption{Distributions of spin ($\lambda^\prime$) for DM halo samples with (blue) and without (orange) a GSE analogue, smoothed with Epanechnikov kernels \citep{Epanechnikov}. The lower panel shows the cumulative distributions, with $f_\mathrm{hosts}(<\lambda^\prime)$ being the fraction of hosts in each sample with a spin parameter less than $\lambda^\prime$. The vertical dashed lines mark the medians of the two populations. Haloes containing a GSE analogue have lower spin parameters on average, with the median $\lambda^\prime$ being a factor of 1.7 smaller. The Kolmogorov-Smirnov (KS) statistic $D$ and $p$-value are printed in the lower panel.}
   \label{fig:spin_dists}
\end{figure}

\section{Simulations}\label{section:simulations}

The \textsc{artemis} suite \citep{Font} consists of zoom-in hydrodynamical simulations of 45 galaxies with MW-mass haloes. A detailed description is given by \citet{Font} and is summarised below.

The simulations were run in a WMAP $\Lambda$CDM cosmology with the Gadget-3 code \citep{Springel_gadget}, using the same hydrodynamics solver and subgrid physics as the EAGLE project \citep{schaye2015,crain2015}. However, the stellar feedback was recalibrated to obtain a better match to the observed stellar mass-halo mass relation \citep[for details see][]{Font}.

The 45 haloes were selected from a collisionless simulation in a box of side length 25~Mpc~$h^{-1}$ with periodic boundary conditions. This selection was purely based on the total mass of the haloes, all of which lie in the range $0.8<M_{200}/10^{12}M_\odot<1.2$. Here $M_{200}$ is the mass enclosed within a volume of mean density 200 times the critical density at redshift $z=0$.

The final simulations have dark matter particles of mass $1.17\times10^5M_\odot h^{-1}$ and baryon particles with initial mass $2.23\times10^4M_\odot h^{-1}$. The Plummer-equivalent softening length is $125$~pc~$h^{-1}$. Haloes and subhaloes were identified with the SUBFIND algorithm \citep{Dolag_subfind}.

The simulated haloes match observed properties of MW-mass galaxies, including stellar masses, sizes, luminosities and average metallicites \citep[see][]{Font}.

\begin{figure}
  \centering
  \includegraphics[width=0.8\columnwidth]{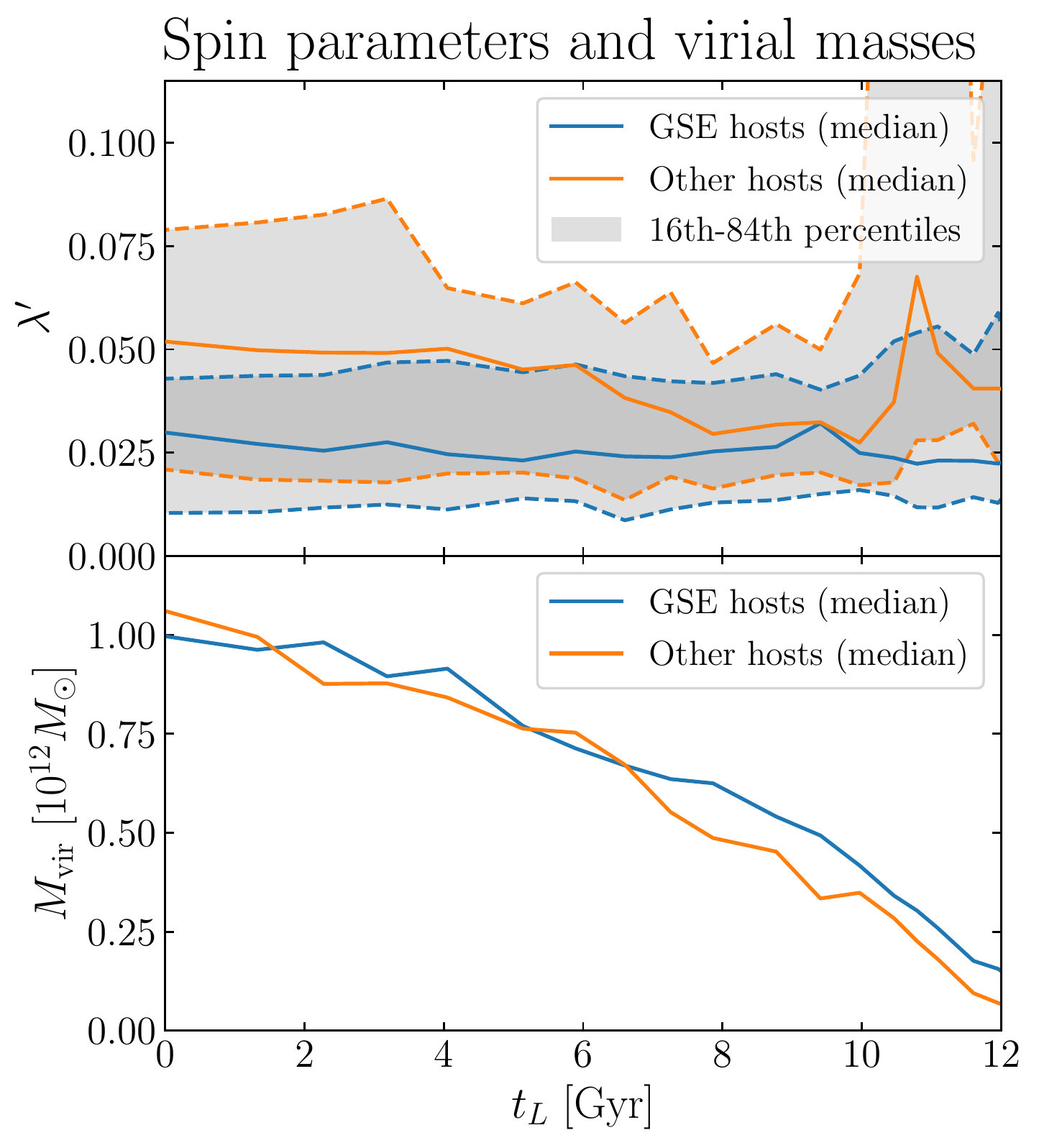}
  \caption{\textbf{Top panel:} Median DM spin parameters (solid lines) of the halo samples with (blue) and without (orange) a GSE analogue as functions of lookback time $t_L$. The grey bands enclosed by the dashed lines indicate the ranges between the 16th and 84th percentiles. At 10 Gyr ago, the medians are similar, before the value for the non-GSE sample increases markedly from 8 to 6 Gyr ago. At the present-day the median lies outside the 84th percentile of the GSE sample. \textbf{Bottom panel}: Median virial mass of each sample as functions of time. The non-GSE haloes initially have lower mass, before growing more rapidly between 8 and 6 Gyr ago to catch up with the GSE sample.}
   \label{fig:spin_mass_tL}
\end{figure}

\begin{figure}
  \centering
  \includegraphics[width=0.95\columnwidth]{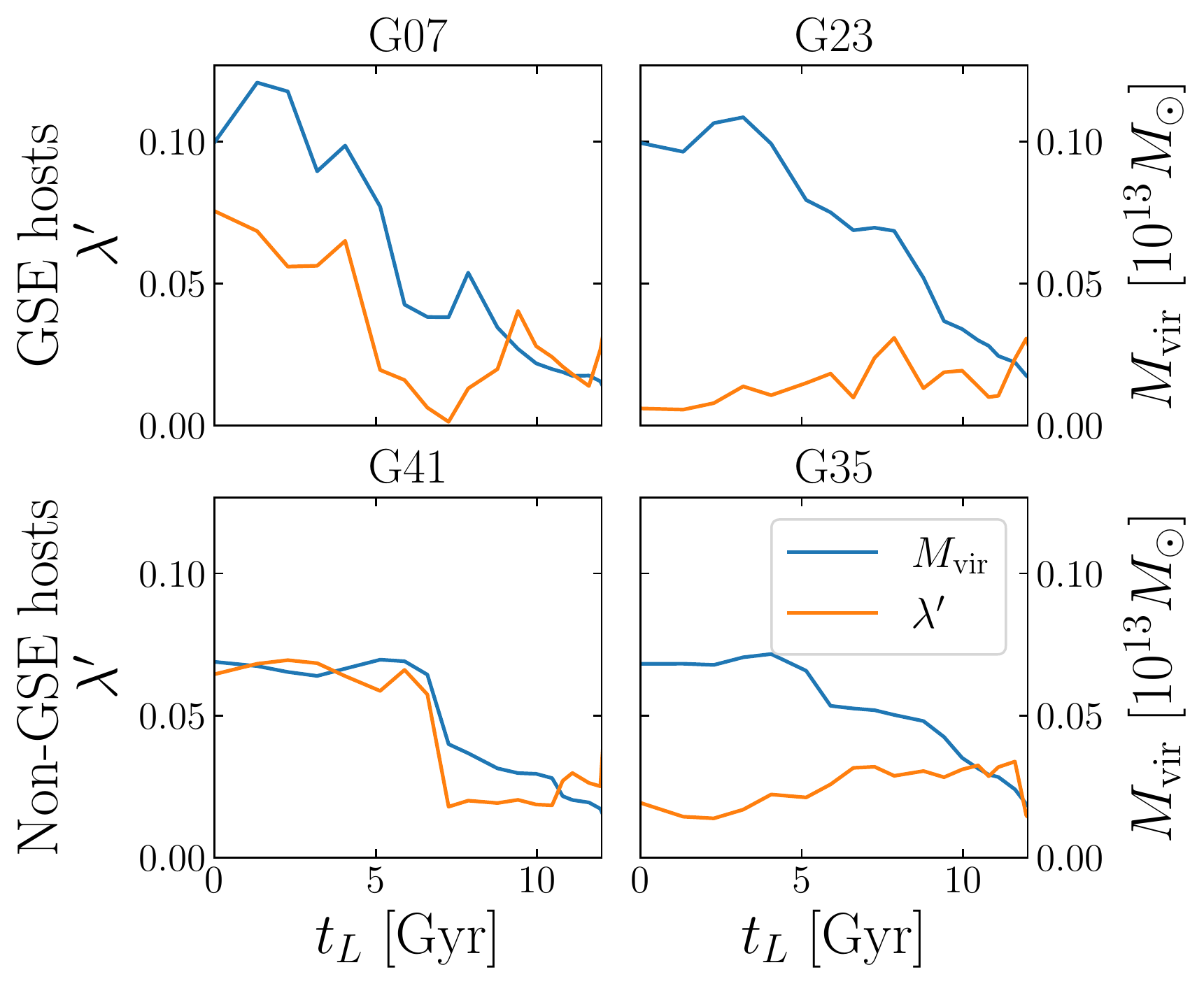}
  \caption{Virial masses (blue) and DM spin parameters (orange) plotted together against time for four haloes. Those in the top row (G07 and G23) belong to the GSE sample, while the others (G41 and G35) do not. In both samples there are clear examples of large increases in virial mass and spin coinciding (e.g. G07 and G41), but also those where $\lambda^\prime$ changes little or decreases when $M_\mathrm{vir}$ rises (G23 and G35).}
   \label{fig:spin_mass_individual}
\end{figure}

\section{Method}\label{section:method}

\subsection{Selection of galaxy samples}

We divide the 45 galaxies into two samples based on the presence of a GSE-like feature in the kinematics of accreted stars. The sample of 23 GSE-containing haloes is taken from \citet{Dillamore} (called Sample GS/all and shown in Fig.~3 in that study), and its selection is described briefly below.

Star particles in the \textsc{artemis} simulations are flagged as either `\textit{in situ}' or `accreted'. \textit{In situ} stars are those born in the host halo, defined as the most massive subhalo in the most massive friends-of-friends group \citep[][]{Font}. Accreted star particles are therefore those which have been stripped from lower mass satellites, usually during merger events.

The velocities of accreted star particles were calculated in spherical coordinates, and a two-component Gaussian mixture model was fitted to the distributions within a Solar neighbourhood region. This is a cylindrical shell between radii $R=5$~kpc and $15$~kpc, between heights of $z=\pm9$~kpc (above and below the galactic plane). The anisotropy parameter $\beta=1-(\sigma^2_\phi+\sigma^2_\theta)/2\sigma^2_r$ was calculated for each Gaussian, where $\sigma_i$ is the velocity dispersion in the $i\mathrm{th}$ direction \citep[see][]{Binney_Tremaine}. The GSE sample consists of all haloes where the more radially anisotropic Gaussian component (larger $\beta$) has a contribution of $>40\%$ and $\beta>0.8$. This procedure is based on that of \citet{Fattahi}, and provides us with a set of galaxies that contain a significant accreted feature with highly radial kinematics.

We emphasise that this selection was purely based on the kinematics of accreted star particles, with no consideration of DM or any other properties of the haloes. Examples of the velocity distributions of accreted stars can be seen in Fig.~1 of \citet{Dillamore}.

\subsection{Calculation of spin parameters}
\label{section:calculation}

Due to the computational difficulties associated with calculating the energy for the spin parameter $\lambda$ (Equation~\ref{equation:lambda}), we follow \citet{Bullock} and instead use the modified spin parameter,
\begin{equation}\label{equation:lambda_prime}
    \lambda' \equiv \frac{J_\mathrm{DM, vir}}{\sqrt{2}M_\mathrm{DM,vir}V_\mathrm{vir}R_\mathrm{vir}}.
\end{equation}
Here, $J_\mathrm{DM, vir}$ is the total angular momentum of DM particles within the virial radius $R_\mathrm{vir}$, of total mass $M_\mathrm{DM,vir}$. The circular velocity at the virial radius is $V_\mathrm{vir}=\sqrt{GM_\mathrm{vir}/R_\mathrm{vir}}$ where $M_\mathrm{vir}$ is the total virial mass.

We use the virial mass and radius values computed by SUBFIND, while the DM angular momentum and mass within $R_\mathrm{vir}$ are calculated using particle data.

\subsection{Testing for equilibrium}

Following \citet{Maccio} and \citet{D'Onghia}, we quantify the haloes' level of equilibrium using the distances between their centres of mass and potential (i.e. location of most bound particle). We define the parameter $s\equiv|\mathbf{r}_\mathrm{CoM}-\mathbf{r}_\mathrm{CoP}|/R_\mathrm{vir}$, where $\mathbf{r}_\mathrm{CoM}$ and $\mathbf{r}_\mathrm{CoP}$ are the position vectors of the centre of mass and potential respectively. The centre of mass is calculated from all particles belonging to the halo, while the centre of potential is the position of its most bound particle \citep{McAlpine}.

Haloes in virial equilibrium tend to have very small values of $s$, while those with large $s$ are unrelaxed, often having undergone recent mergers \citep{Maccio}. This quantity is therefore a useful proxy for measuring the virial equilibrium of a halo, and is computationally easy to evaluate.

\begin{figure}
  \centering
  \includegraphics[width=\columnwidth]{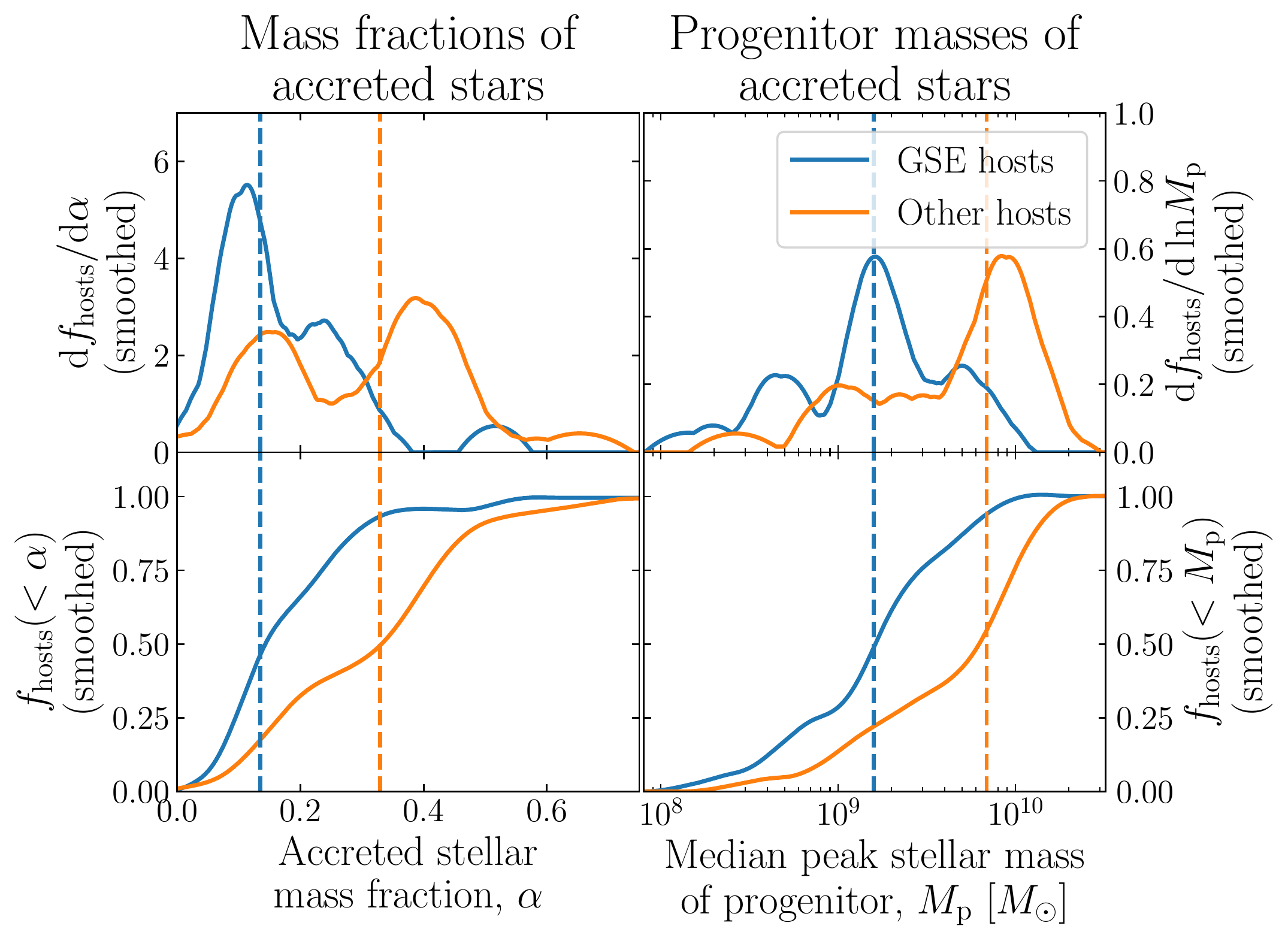}
  \caption{\textbf{Left column:} Differential (top) and cumulative (bottom) distributions of the mass fractions of stars which are flagged as accreted \citep[smoothed with Epanechnikov kernels,][]{Epanechnikov}. $f_\mathrm{hosts}(<\alpha)$ is the fraction of hosts with mass fraction less than $\alpha$. The medians are marked by dashed lines. The haloes with no GSE analogue have larger fractions of accreted stars by more than a factor of 2 on average. \textbf{Right column:} Distributions of the median peak stellar mass of the progenitors to which the accreted stars belonged before they were stripped. This is a measure of the typical masses of accreted satellites. The stars in the non-GSE sample tend to originate in galaxies with higher stellar mass, on the order of $10^{10}M_\odot$ compared to $10^9M_\odot$ for the GSE sample.}
   \label{fig:accreted_fraction_dists}
\end{figure}

\section{Results}\label{section:results}

\subsection{DM spin distributions}

The present-day distributions of $\lambda'$ are shown in Fig.~\ref{fig:spin_dists} for the halo samples with and without a GSE analogue. The two distributions appear to be distinct, with the spin of the GSE DM haloes being smaller on average; the median values of $\lambda'$ differ by a factor of about 1.7. The results of a two-sample Kolmogorov-Smirnov (KS) test are displayed in the lower panel. The KS statistic $D$ is the maximum vertical distance between the cumulative distributions, while the $p$-value is the probability that the two samples are drawn from the same probability distribution. The reasonably small $p$-value of 0.012 supports the hypothesis that these distributions are truly distinct.

We must address whether this distinction in spin distributions is due to the non-GSE hosts being further from equilibrium, as suggested by \citet{D'Onghia}. The off-centre parameters $s$ are indeed larger on average for the non-GSE hosts. The median value is $s=0.055$, compared to $0.036$ for the GSE hosts. The haloes without a GSE analogue therefore typically have a larger offset between their centres of mass and potential, implying that they are on average further from virial equilibrium. This suggests that a lack of equilibrium may be a factor causing the higher spins of the non-GSE hosts.

Alternatively, we can check whether cutting the samples based on $s$ affects the spin distributions. We remove all haloes from both samples with $s>0.1$, the upper threshold for virial equilibrium used by \citet{D'Onghia}. This removes 3 haloes from the GSE sample and 7 from the non-GSE sample. After this cut the median values of $\lambda^\prime$ are 0.024 and 0.046 for the GSE and non-GSE samples respectively, a difference of a factor of 1.9. Hence the median values have actually moved further apart. The KS statistic is now 0.43 and the $p$-value is 0.061 (the larger $p$-value is partly due to the smaller sample sizes). Though the probability of distinct distributions is slightly reduced, there remains a clear difference between the average spins of the two samples when the out-of-equilibrium haloes are removed. This suggests that lack of virialization alone may not be the cause of the distinction.

In the top panel of Fig.~\ref{fig:spin_mass_tL}, we show the evolution of the spin parameters across lookback time $t_L$. The medians and 16th and 84th percentiles of $\lambda^\prime$ are plotted for each sample. The bottom panel shows the median virial masses $M_\mathrm{vir}$.

The two distributions are clearly distinct at $t_L\approx11$~Gyr, with some non-GSE hosts exhibiting very large values of $\lambda^\prime$. However, during this time the haloes are accreting mass rapidly, so may be far from virial equilibrium. This may explain the high spins; \citet{D'Onghia} found that out-of-equilibrium haloes have higher spins on average. The virial radius and mass is also more difficult to define if a halo is rapidly accreting and far from equilibrium, so $\lambda^\prime$ (as defined by Equation~\ref{equation:lambda_prime}) may not be meaningful.

At $t_L\approx9-10$~Gyr, the two $\lambda^\prime$ distributions are more alike than at present, with similar median values. They then diverge over the next several billion years, with the non-GSE median increasing from $\lambda^\prime\approx0.03$ to $\approx0.05$. Much of this change (particularly in the median) happens between 8 and 6 Gyr before present. This period coincides with an interval during which the non-GSE hosts experience particularly rapid mass growth. Until $t_L\approx8$~Gyr, the GSE hosts have larger virial masses on average, with the others lagging behind. This changes over the next 2 Gyr, when the median virial mass of the non-GSE hosts increases significantly and overtakes that of the GSE hosts. The two averages then remain relatively close until the present.

\begin{figure}
  \centering
  \includegraphics[width=0.9\columnwidth]{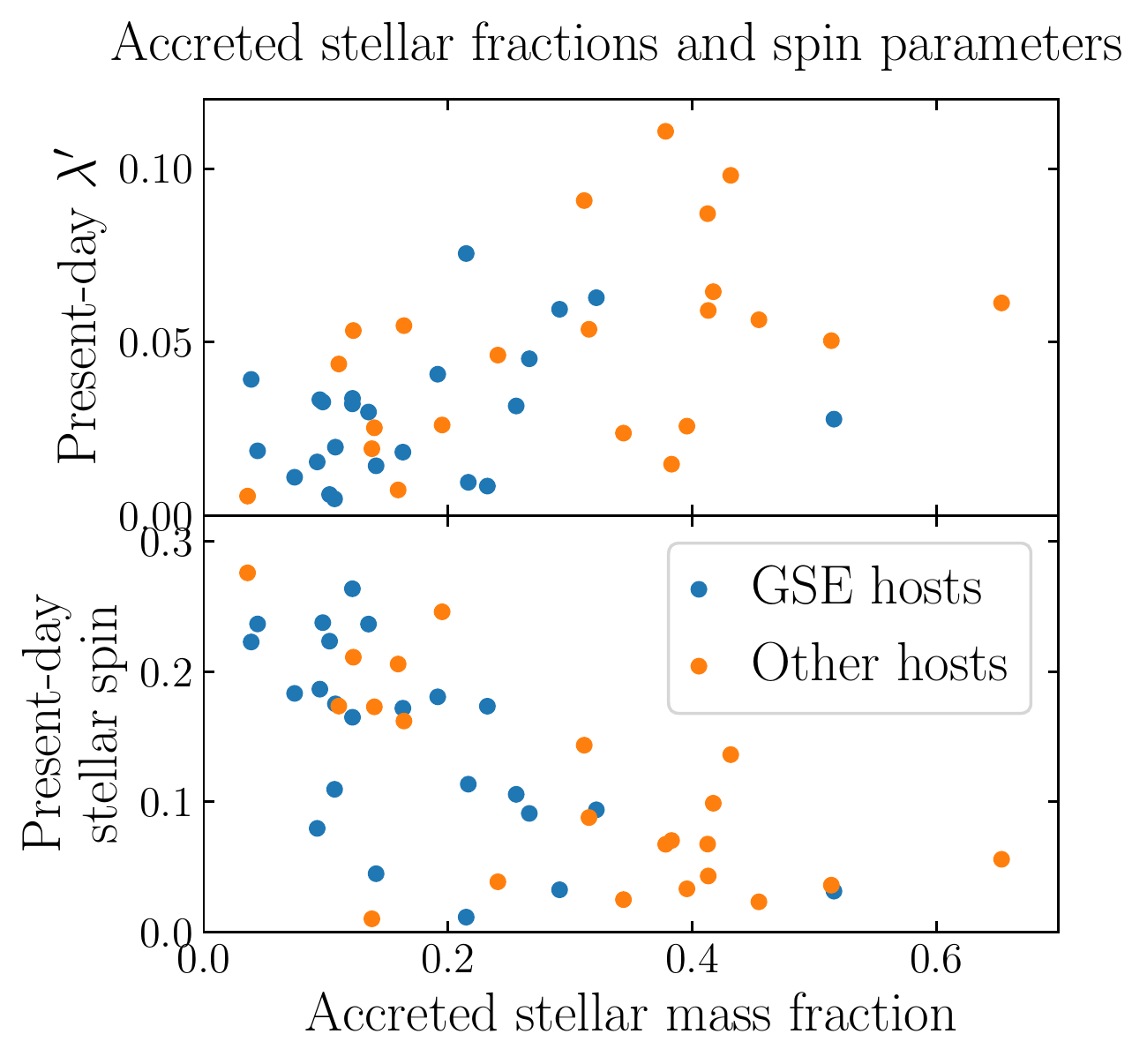}
  \caption{\textbf{Top panel:} Present-day DM spin parameter $\lambda^\prime$ against accreted stellar mass fraction, for both halo samples. Higher accreted fractions allow a wider range of $\lambda^\prime$, with $\lambda^\prime\gtrsim0.6$ only occurring when the accreted fraction is greater than 0.2. \textbf{Bottom panel}: As above, but stellar spin instead of DM halo spin. In this case the correlation is negative, with larger proportions of accreted stars corresponding to lower spins.}
   \label{fig:accreted_fraction_spin}
\end{figure}

\subsection{Accretion histories}

The concurrence of the increases in spin and virial mass of the non-GSE hosts hints that mergers may play an important role in the spin evolution \citep[e.g.][]{Gardner, Vitvitska, Peirani, D'Onghia}. In Fig.~\ref{fig:spin_mass_individual}, we show the spin and virial mass evolution for two individual galaxies in each sample. The two haloes in the left-hand column (G07 and G41) show close correspondence between the virial mass and spin. During major mergers when the virial mass increases rapidly, the spin also increases sharply over the same period. In both of these cases, the spin remains high after the merger has concluded. The stability of the spin (for 6~Gyr in G41) suggests that the increase is not merely due to a lack of virial equilibirum, but is truly driven up by the accretion. However, this is not universally the case. Some haloes (including G23 and G35) do not experience large spin increases during major accretion episodes. Both galaxy samples exhibit examples of both behaviours. However, the fact that high spins tend to arise after major accretion events suggests that a difference in accretion history may be a dominant factor in the distinct present-day spin distributions.

If this is the case, it is expected that the populations of accreted stars will also show differences. Properties of stellar populations are more directly observable than those of the DM halo, so any relation here is more readily applicable to the MW. In Fig.~\ref{fig:accreted_fraction_dists} we show distributions of two properties of the accreted stellar populations. Firstly we calculate the accreted stellar mass fraction (that is, total mass of accreted stars divided by total stellar mass). This is shown in the left-hand column, with the cumulative distribution in the lower panel. For each star, we also consider the peak stellar mass of any progenitor halo to which it was bound before it was accreted onto the host. The median of this quantity across all accreted stars is calculated for each galaxy, and the distributions are plotted in the right-hand column of Fig.~\ref{fig:accreted_fraction_dists}. This is a measure of the typical stellar mass of satellites accreted by the hosts.

The accreted stellar mass fraction tends to be much greater in the non-GSE hosts, with a median value of 0.33 compared to 0.13 in the GSE hosts. The right-hand column shows that the stars accreted onto the non-GSE hosts tend to originate in much higher mass progenitors; accreted stars in GSE hosts mostly come from haloes of stellar mass $\lesssim3\times10^9M_\odot$. By comparison, the non-GSE hosts accrete satellites with typical masses of $\sim7\times10^{9}M_\odot$ and up to more than $10^{10}M_\odot$. The \textsc{artemis} hosts themselves have present-day stellar masses of typically $2-4\times10^{10}M_\odot$ \citep{Font}, so some of these satellites are only slightly smaller than the hosts. The non-GSE accreted fractions are also significantly larger than that of the MW; the MW's stellar halo makes up only about $2\%$ of its total stellar mass \citep{Licquia,Deason_halomass}, so its accreted fraction is even less than this. It appears that a lack of a GSE analogue tends to indicate higher mass mergers and accreted fractions. We have found that accreted stellar haloes in the non-GSE sample tend to have lower radial anisotropy across all redshifts, and experience mergers with mass ratios of typically $\sim0.5$. We therefore postulate that mergers with such a high mass ratio do not give rise to a GSE analogue, instead resulting in a more isotropic accreted velocity distribution.

The correlation between large numbers of accreted stars and massive accreted haloes is unsurprising. Less massive (dwarf) galaxies are dark matter dominated \citep[e.g.][]{Moster}, so many minor mergers with these would not easily provide large quantities of accreted stars to the host haloes. More massive haloes have proportionally much larger stellar masses, so a merger with one of these could provide the large accreted stellar fractions seen in the non-GSE sample. For example, a typical \textsc{artemis} stellar mass of $2\times10^{10}M_\odot$ and accreted stellar fraction of 0.35 could correspond to a single merger with a satellite of stellar mass $7\times10^9M_\odot$, which is roughly average for the non-GSE sample.

In the top panel of Fig.~\ref{fig:accreted_fraction_spin} we plot the halo spin parameters against the accreted stellar mass fractions. This again shows that the GSE hosts have lower spin parameters and accreted fractions. It also emphasises that high spin parameters ($\gtrsim0.6$) only occur along with high accreted mass fractions ($\gtrsim0.2$), although there is a wide range of $\lambda^\prime$ at high accreted fractions.

In the bottom panel we instead show the spins of the stellar components of the galaxies. We define this in the same way as $\lambda^\prime$ (Equation~\ref{equation:lambda_prime}), but replace the DM angular momentum and mass with those of the stars, and evaluate all quantities at the galaxy's effective radius (i.e. half-mass radius). This plot shows the opposite trend: galaxies with higher fractions of accreted stars have lower net stellar spin. This is unsurprising for two reasons: massive mergers are able to disrupt stellar discs \citep[see e.g.][]{Hopkins}, and a larger proportion of accreted stars means a smaller proportion of \textit{in situ} stars, many of which occupy a disc and therefore have closely correlated angular momenta.

\section{Summary}\label{section:summary}

We have measured the spin distributions of MW-mass galaxies in the \textsc{artemis} cosmological simulations, dividing the 45 haloes according to the presence or absence of an analogue of \textit{Gaia} Sausage-Enceladus (GSE). A GSE analogue is defined as a set of accreted stars on highly radial orbits, which has a large contribution to the overall accreted population \citep[as described by][]{Dillamore}. We find that:

\begin{enumerate}[label=\textbf{(\roman*)}]
    \item The haloes containing a GSE analogue have lower spin parameters on average. The median dimensionless spin parameter $\lambda^\prime$ is 0.030 for the GSE sample, and 0.052 for the others (a difference of a factor of 1.7). The $p$-value of a two-sample KS test is 0.012, supporting the conclusion that the spin distributions of the two samples are distinct.
    
    \item At earlier times (8-10 Gyr before present) the spin distributions of the two samples are much more similar. They start to diverge after $t_L\approx8$~Gyr, when the median $\lambda^\prime$ of the non-GSE hosts increases from $\approx0.03$ to $\approx0.05$. This coincides with an increased rate of growth of the median virial mass of this sample.
    
    \item The GSE hosts have smaller fractions of accreted stars than the other hosts. A median of $13\%$ of the stellar mass in GSE hosts is accreted from other subhaloes, compared to $33\%$ for the non-GSE hosts. This accreted stellar mass also tends to come from smaller satellites, typically $\sim10^9M_\odot$ for the GSE hosts compared to $\sim0.7\times10^{10}M_\odot$ for the others.
    
    \item High present-day DM spin parameters ($\lambda\gtrsim0.06$) only occur when the mass fraction of accreted stars is $\gtrsim0.2$, suggesting that mergers may be necessary for driving large spins.
\end{enumerate}

Our results demonstrate the importance of a galaxy's stellar component in the study of its DM halo. If we wish to constrain the spin of the Milky Way's dark halo, consideration of the stellar halo and accretion history may be crucial.

\section*{Acknowledgements}

We are grateful to Ian McCarthy for providing supplementary simulation data used in this study. We also thank the anonymous referee for helpful comments used to improve this manuscript.

AMD thanks the Science and Technology Facilities Council (STFC) for a PhD studentship.
This work was funded by UKRI grant 2604986. For the purpose of open access, the author has applied a Creative Commons Attribution (CC BY) licence to any Author Accepted Manuscript version arising.

\section*{Data Availability}

Data from the \textsc{artemis} simulations may be shared on reasonable request to the corresponding author.



\bibliographystyle{mnras}
\bibliography{bibliography} 




\appendix


\bsp	
\label{lastpage}
\end{document}